\documentclass[11pt,a4paper]{article}
\pdfoutput=1
\usepackage{lscape}
\usepackage{caption}
\usepackage{jheppub}
\usepackage[utf8]{inputenc}
\usepackage{soul,xcolor}
\usepackage{cancel}
\usepackage{makecell}
\usepackage{diagbox}
\usepackage[normalem]{ulem}
\newsavebox{\foobox}

\usepackage{multirow}
\usepackage{dcolumn}
   
\usepackage{ulem}
\usepackage{graphicx}
\usepackage{comment}
\usepackage{bm,array}
\usepackage{graphics}
\usepackage{epsf}
\usepackage{color}
\usepackage{amsmath}
\usepackage{amssymb}
\usepackage{latexsym}
\usepackage{slashed}
\usepackage{float}
\usepackage{cases}
\usepackage{multirow}
\usepackage{framed}
\def\be{\begin{equation}}
\def\ee{\end{equation}}
\def\ba{\begin{array}}
\def\ea{\end{array}}
\usepackage{amsmath}
\usepackage{amssymb}
\usepackage[utf8]{inputenc}
\def\sQ3{\widetilde{Q}_3}
\def\sU3{\widetilde{U}_3}
\def\sD3{\widetilde{D}_3}

\def\hsm{h_{\rm SM}}

\def\d{\partial}

\def\mhsm{m_{h_{\mathrm{SM}}}}

\usepackage{graphicx}
\usepackage{tikz}
\usetikzlibrary{snakes}
\usepackage{physics}

\def\vev{{\it vev}}

\def\beq{\begin{equation}}
\def\eeq{\end{equation}}
\def\beqa{\begin{eqnarray}}
\def\eeqa{\end{eqnarray}}

\allowdisplaybreaks

\def\micromegas{{\tt micrOMEGAs}}
\def\feynrules{{\tt FeynRules}}

\def\pythia8{{\tt PYTHIA8}}

\def\z3nmssm{$Z_3$-NMSSM}
\newcommand{\bea}{\begin{eqnarray}}
\newcommand{\eea}{\end{eqnarray}}

\title{{\Large Dilution of Dark Matter Relic abundance due to First Order Electroweak Phase Transition}}
\author[a,b,c]{Subhojit Roy}
\affiliation[a]{Harish-Chandra Research Institute, A CI of Homi Bhabha National
Institute, Chhatnag Road, Jhunsi, Prayagraj (Allahabad) 211019, India}
\affiliation[b]{Regional Centre for Accelerator-based Particle Physics, Harish-Chandra Research Institute, \\ Prayagraj (Allahabad) 211019, India}
\affiliation[c]{HEP Division, Argonne National Laboratory, 9700 Cass Ave., Argonne, IL 60439, USA}
\emailAdd{subhojitroy@hri.res.in, sroy@anl.gov}
\preprint{HRI-RECAPP-2022-015}
\abstract{We investigate the effect of a first-order electroweak phase transition~(FOEWPT), which is one of the prerequisites for electroweak baryogenesis,  on the thermal relic abundance of the dark matter (DM) that freezes out before the occurrence of the phase transition in the complex singlet scalar extended $Z_3$-invariant type-II seesaw model that can simultaneously provide a DM candidate, explain the non-vanishing neutrino masses and the baryon asymmetry of the Universe. 
Such a phase transition around the electroweak scale leaves an impact on the relic density due to the release of entropy, particularly for a TeV-scale DM.
We thus concentrate on the region of parameter space of the said model, which favors an FOEWPT in the early Universe and for which the DM is heavy such that its freeze-out temperature turns out to be larger than the phase transition temperature.   
We further study the dependencies of the dilution factor of the DM relic density on the model parameters, the nucleation temperature, the strength and the duration of the phase transition. Such a dilution might retrieve some of the regions of parameter space that were previously ruled out
by the measured value of the DM relic density and/or the latest constraints from the DM direct-detection experiments.
Furthermore, a direct connection is drawn between the dilution factor and the generation of stochastic gravitational waves as a result of an FOEWPT.
}
\begin{document}
\maketitle
%
\section{Introduction}
\label{Introduction}

The relic density of dark matter (DM) in the present Universe is precisely measured from various astrophysical and cosmological experiments, i.e., $\Omega_{\rm DM} h^2 = 0.120 \pm 0.001$~\cite{Planck:2018vyg}.
The particle responsible for the observed DM in the Universe is likely
the most wanted new particle after the latest finding of the Standard Model (SM)-like Higgs boson mass
around 125 GeV.
Among the various possibilities of the nature of the DM, the weakly interacting massive particle (WIMP) still remains a rather viable and the well-motivated DM candidate despite null results from numerous DM experiments and various searches at the Large Hadron Collider (LHC). 

The discovery of the Higgs boson~\cite{ATLAS:2012yve, CMS:2012qbp} at the LHC completes the 
findings of all the particles appearing in the
SM particle physics which vindicates the scheme of spontaneous electroweak symmetry breaking (SSB) as incorporated in the SM.
Nevertheless, the electroweak symmetry is restored in the early Universe, i.e., at high temperatures~\cite{Kirzhnits:1972ut}. It is generally understood that SSB transition occurs in the SM via a smooth cross-over~\cite{Kajantie:1996qd}. However, in the presence of additional scalars in a scenario beyond the SM (BSM), 
a first-order phase transition (FOPT) could occur.
FOPT at the electroweak scale has an advantage in that it can provide one of the required conditions to explain the baryon asymmetry of the Universe (BAU).
The conventional measure of BAU, 
$Y_B$, is the ratio of the difference in the number density of baryon and anti-baryon
($n_B$ and $n_{\bar{B}}$, respectively) and the entropy density ($s$), i.e., 
$Y_B=(n_B-n_{\bar{B}}) / s$.
The most accurate value of $Y_B$ ($=8.65 \pm 0.09 \times 10-11$) comes from the Planck experiment~\cite{Planck:2015fie} from its
 measurement of the baryon acoustic oscillations in the power spectrum of the cosmic microwave background.

In order for baryogenesis to take place, the following three Sakharov requirements~\cite{Sakharov:1967dj} must be met: 
(i) non-conservation of baryon number ($\slashed{B}$), (ii) $C$ and $CP$ violations
($\slashed{C}$, $\cancel{CP}$) and (iii) departure from thermal equilibrium.
Electroweak baryogenesis (EWBG) is one of the most attractive mechanisms to explain the BAU. It has been widely studied in the literature~\cite{Trodden:1998ym, Anderson:1991zb, Huet:1995sh, Morrissey:2012db} and it requires an FOPT in the electroweak Higgs sector. FOPT fulfils the third criterion of Sakharov's conditions, i.e., the system has to go out of thermal equilibrium. In addition, the phase transition has to be strongly first-order to avoid a wash-out of the created baryons inside the broken EW phase~\cite{Quiros:1999jp}. 

The study of FOPT at around the electroweak scale has recently gained heightened attention, particularly after the recent observation of gravitational waves (GW) from LIGO and VIRGO collaboration~\cite{LIGOScientific:2016aoc, LIGOScientific:2017vwq, LIGOScientific:2018mvr, LIGOScientific:2020ibl}. This is since FOPT would be responsible for possible stochastic GW which might act as a background for the various future space/ground-based GW experiments, viz.,
LISA~\cite{LISA:2017pwj}, ALIA~\cite{Gong:2014mca}, Taiji~\cite{Hu:2017mde}, TianQin~\cite{TianQin:2015yph}, aLigo+~\cite{Harry:2010zz}, Big Bang Observer (BBO)~\cite{Corbin:2005ny} and  Ultimate(U)-DECIGO~\cite{Kudoh:2005as}. Note that EWPT in the SM is unable to generate any GW since the transition is of a cross-over type. Therefore, detection of such type of a stochastic GW would likely point to BSM physics.

In addition, an FOPT in the early Universe can significantly impact the thermal history of those particle(s) species that freeze out before the phase transition. During the phase transition of the Universe from the false to true vacuum, the relic density of those species is diluted due to an injection of entropy and a release of latent heat. Today's relic abundance of the thermal species that decouple after the phase transition would not be affected as a thermal equilibrium attained after the phase transition would erase the effect of dilution. 

The effect of dilution on the DM relic density would be significant when the temperature of the FOPT happens to be smaller than the DM freeze-out temperature. Such an effect would therefore be important in the context of a first-order electroweak phase transition (FOEWPT) for DM candidates with TeV-scale masses ($m_{\chi}$) which are expected to get frozen out from the thermal equilibrium at a freeze-out temperature  ($T_{\rm FO}$) around the electroweak scale ($T_{\rm FO} \backsimeq \frac{m_{\chi}}{25}$).
The dilution of thermal relic abundance of DM due to FOPT has been studied in the past by considering some toy-model-like effective potential such as adding a larger degree of freedom from the hidden sector~\cite{Megevand:2003tg, Megevand:2007sv,Wainwright:2009mq,Chung:2011hv}. Recently, it has been studied in more realistic models such as a singlet extension of the SM (xSM) and two-Higgs-doublet model (2HDM+S) in reference~\cite{Xiao:2022oaq}.

Apart from the issues pertaining to DM and BAU, another shortcoming of the SM is that it considers the neutrinos to be massless while various experiments have established that neutrinos indeed have mass~\cite{Super-Kamiokande:1998kpq, SNO:2001kpb}.
In this work, we consider $Z_3$-symmetric complex singlet scalar extension type-II seesaw model which 
has a natural explanation of the origin of finite neutrino mass, provides a singlet scalar DM candidate and can generate the observed baryon asymmetry of the Universe via the EWBG.
In contrast to the type-I and type-III seesaw models, the type-II seesaw model permits large neutrino Yukawa couplings via a small triplet vacuum expectation value concurrently even with a sub-TeV seesaw. In addition, new interactions between the SM Higgs doublet and the complex triplet could modify the Higgs potential in favor of an FOPT.
The interplay among the production of GW,
DM and physics at the LHC has been studied in this model in reference~\cite{Ghosh:2022fzp}.
Recently, it has been shown that this model can explain the observed excess of electron-positron flux from AMS-02~\cite{AMS:2014xys}, DAMPE~\cite{DAMPE:2017fbg} and Fermi-LAT~\cite{Fermi-LAT:2017bpc} collaborations~\cite{Yang:2021buu}.
The present work aims to investigate a possible dilution of the relic abundance of a heavy
 singlet-like scalar DM due to an FOPT in the Higgs sector around the electroweak scale.
We calculate the magnitude of the dilution factor and point out the region of the relevant  
parameter space where such a dilution could be significant. 
Furthermore, a direct connection between the dilution factor and the generation of stochastic GW as a result of an FOEWPT is also discussed.

In this work, we focus on the dilution of the DM relic abundance due to an FOPT occurring around the electroweak scale, which corresponds to a specific region of the model’s parameter space.
As mentioned earlier, such dilution primarily affects DM particles whose freeze-out temperatures are larger than the phase transition’s completion temperature, approximately the so-called nucleation temperature.
This implies that DM must be relatively heavy ($\gtrsim 1$~TeV) if the first-order phase transition (FOPT) occurs near the electroweak scale ($\sim 100$~GeV), ensuring that its freeze-out temperature exceeds the nucleation temperature. In such a scenario, the dark sector experiences significant Boltzmann suppression, characterized by $e^{-m/T}$, and remains effectively decoupled from the Higgs sector during the phase transition. As a result, the dark sector does not influence the FOPT of interest around the electroweak scale in this study. Instead, the FOPT is mostly driven by the Higgs sector which is consist with a Higgs triplet and a Higgs doublet.

The complex scalar singlet models featuring a $Z_3$ symmetry have been extensively explored in the literature. For instance, Refs.~\cite{Ko:2014nha,Bernal:2015bla} investigate Strongly Interacting Massive Particles (SIMPs) by introducing a non-zero vacuum expectation value (VEV) for a dark Higgs field under $Z_3$ symmetry, yielding DM candidates within the mass range ${\cal O}(1-100)$ MeV. Ref.~\cite{Hektor:2019ote}, on the other hand, refines the constraints on $Z_3$ singlet DM by incorporating updated unitarity bounds and a detailed analysis of early kinetic decoupling. These studies reveal two allowed DM mass ranges: $(56.8\sim 58.4)\ {\rm GeV} \leq m_{\rm DM} \leq 62.8\ {\rm GeV}$ and $m_{\rm DM} \geq 122\ {\rm GeV}$, particularly when semi-annihilation processes are significant during freeze-out.
In this work, the $Z_3$-symmetry of the model enhances the dark sector by introducing a semi-annihilation channel that influences the relic density. The impact of this semi-annihilation process has been thoroughly analyzed in Refs.~\cite{Yang:2021buu, Ghosh:2022fzp}~\footnote{An intriguing phenomenological feature of the $Z_3$ discrete symmetry is that it introduces cubic interactions in the dark sector, which can create a barrier in the scalar potential at finite temperature, thereby triggering an FOPT~\cite{Ghosh:2022fzp}. These FOPTs in the dark sector can generate GWs, increasing the potential for testing certain parameter space of such models in future proposed GWs experiments.
}. Although the FOEWPT responsible for diluting the relic density is not related to the dark sector, the role of the discrete symmetry here is minimal, as it is only necessary to stabilize the dark matter. Replacing the $Z_3$ symmetry with a $Z_2$ symmetry would modify the allowed parameter space of the dark sector, but would not impact the dilution factor calculated in this study, since the dilution factor depends solely on the Higgs triplet and the SM-like Higgs doublet sector. Note that, in our previous work~\cite{Ghosh:2022fzp}, we investigated the interplay among dark sector phenomenology, gravitational wave production from FOPTs, and collider searches. Building on that foundation, this study extends our analysis to explore the dilution effect of FOPTs on the relic abundance within a specific region of the model’s parameter space.

The present work is organized as follows. We briefly review how an FOPT affects the relic abundance of DM particles that freeze out later than the transition temperature in section~\ref{dilution}.  The detailed description of the model is discussed in section~\ref{model}. We present our results in section~\ref{result}. In section~\ref{conclusions}, we conclude. The estimation of the production of stochastic GW from sound waves is discussed in the appendix~\ref{GWexpression}.
\section{Dilution of DM relic density due to FOPT}
\label{dilution}
In this section, we discuss the dilution of thermal DM due to the release of entropy during an FOPT that occurs at a considerably lower temperature than $T_{\rm FO}$. The dilution factor depends critically on the temperature at which FOPT occurs. Thus, we will start off by talking about the study of phase transition of the effective scalar potential at finite temperatures. We will later go over how to estimate the dilution factor due to the release of entropy during the FOPT.

When the finite-temperature effective Higgs potential, $V(\phi, T)$, exhibits (at least) two local minima that are already separated by a barrier, an FOEWPT can start to occur in the early Universe. In the hot and radiation-dominated era of the early Universe at high temperatures, the minimum of the scalar potential is at the origin in the field space ($\phi = 0$), i.e., the electroweak symmetry is unbroken.
As the temperature drops due to the expansion of the Universe, a second minimum away from the origin in the field value ($\phi = \phi_{\text{broken}}$) may develop at which $V(\phi_{\text{broken}}, T) > V(0, T)$.
Note that the value of $\phi_{\text{broken}}$ varies with temperature.
With the drop in the temperature of the Universe, $V(\phi_{\text{broken}}, T)$  starts to decrease relative to $V(0, T)$ and at some temperature $T=T_c$, where $T_c$ is known as the `critical temperature of the phase transition, both the two minima become degenerate. 
As the temperature continues to drop further below $T_c$, the minimum at $\phi = \phi_{\text{broken}}$ becomes the global minimum of the effective potential while the minimum at the origin is now considered to be a false minimum.
When there is a barrier between the two minima, the system can start to tunnel from the false minimum to the true global minimum of the potential. 
As a result, at a certain temperature $T<T_c$, a phase transition like this one, known as a first-order type phase transition, can take place in the early Universe.

The model of bubble nucleation is an effective way to describe this type of tunneling process. 
Because of the large surface tension of the bubble walls compared with the energy difference between the two minima, most of the bubbles of the broken minimum remain small and finally collapse.
The bubble of the broken phase nucleates in the cosmological plasma, in which the electroweak symmetry is unbroken, when the rate of nucleation per unit volume ($\Gamma$) is larger than the Hubble expansion rate.
A bubble that is already formed in the plasma continues to develop, collide, and coalesce with other bubbles which are also forming until a big bubble is created. In this manner, the whole space is engulfed and EWSB spreads across the entire field space. 
 At finite temperatures ($T$), in the semi-classical approximation
the tunneling probability from the false vacuum to the true vacuum per unit time per unit volume is given by~\cite{Turner:1992tz}
\beq
\Gamma(T)\simeq T^4 \left( \frac{S_3\left(T \right)}{2\pi T} \right)^{3/2}e^{-S_3\left(T \right)/T},
\eeq
where the three-dimensional Euclidean action $S_3\left(T \right)$ of the background field ($\phi$), in the spherical polar coordinate, is given by
\begin{equation}\label{eq:64}
S_3\left(T \right)=4\pi \int dr \hspace{1mm}r^{2} \left[ \dfrac{1}{2} \left(\partial_{r} \vec{\phi} \right)^2 +V\right]\,\,.
\end{equation}
By extending this Euclidean action, it is possible to determine the prerequisite of the formation of a bubble of a critical size.
In this work, we consider the publicly accessible package $\mathbf{CosmoTransitions}$~\cite{Wainwright:2011kj} for this investigation and to calculate the bounce solution for the aforementioned Euclidean action.
The nucleation temperature can be found by solving the following equation:
\beq
\int_{T_n}^{\infty}\frac{dT}{T}\frac{\Gamma(T)}{H(T)^4} \simeq 1. 
\eeq
The above relation indicates that the probability of a single bubble nucleating inside one horizon volume is around 1. 
It estimates the required condition $S_3(T)/T \approx 140$ to obtain the nucleation temperature ($T_n$) of the phase transition~\cite{Apreda:2001us}. $T_n$ is the highest temperature at which $S_3/ T \lesssim 140$. 
The absolute values of $T_c$, $T_n$ and their difference ($\Delta T_{cn} = T_c - T_n$) have a significant impact on the dilution factor of the relic density of DM due to FOPT.

Depending on the value of $\Delta T_{cn}$, the study of dilution of DM relic density due to the release of entropy caused by an FOPT can be classified into two cases. First, consider the case where $\Delta T_{cn}$ is small. Here, the temperature at which the phase transition occurs is close to the critical temperature of the effective potential which corresponds to negligible supercooling~\cite{Megevand:2007sv,Anderson:1991zb}. In this scenario, the total entropy is conserved as the system is almost in the equilibrium, i.e., $sa^3$ is conserved where `$s$' denotes the entropy density and `$a$' represents the scale factor of the expansion of the Universe. Therefore the entropy density of a phase can be found using this conservation relation. In the following discussion, $+$ ($-$) indicates the high-temperature symmetric (low-temperature broken) phase and the subscript $i$ ($f$) denotes the beginning (end) of the FOPT. Due to the release of entropy during an FOPT, the entropy density changes as
\beq
\label{rntpyconvcase1}
s_{-} = \bigg(\frac{a_i}{a_f}\bigg)^3 s_{+} \, \,.
\eeq
Thus, in this scenario, the dilution factor due to an FOPT is given by
\beq
\label{dilutioncase1}
d \equiv \bigg(\frac{a_f}{a_i}\bigg)^3 = \bigg(\frac{s_{+}(T_c)}{s_{-}(T_n)}\bigg) \, \, .
\eeq
The second case corresponds to the larger $\Delta T_{cn}$ scenario, i.e., when $T_n << T_c$. Equation~\ref{rntpyconvcase1} does not hold as the symmetric and the broken phases are no longer in equilibrium at the beginning of the phase transition. The evolution of the phase transition in this category can be described in three distinct stages: supercooling, reheating, and phase coexistence. The high-temperature phase dominates the Universe during the supercooling stage, wherein the entropy remains conserved. Therefore, similar to equation~\ref{rntpyconvcase1} the following relation can be found
\beq
\label{dilutioncase1}
\bigg(\frac{a_i}{a_s}\bigg)^3 = \bigg(\frac{s_{+}(T_n)}{s_{+}(T_c)}\bigg) \, \,,
\eeq
where $a_s$ is the scale factor at the end of the supercooling stage, i.e., near $T \simeq T_n$.
At the end of the supercooling stage, the release of the latent heat reheats the Universe. The overall energy density of the Universe, $\rho$, does not change if the reheating proceeds far faster than the Universe expansion rate.
However, in this situation, entropy does not continue to be preserved. The Universe can reach the stage of phase coexistence at $T= T_c$ if a significant amount of reheating takes place. 
Its energy density would then be given by
\beq
\label{enrgyconv2ndcase}
\rho_{-}(T_c) = f\rho_{-}(T_c) + (1-f) \rho_{+}(T_c),
\eeq
where $f$ is the volume fraction of the plasma of the low-temperature phase at the beginning of the phase coexistence which can be found from the energy conservation relation given by
\beq
\label{energyconvcond2}
\rho_{+}(T_n) = \rho_{-} (T_c) \, .
\eeq
Using equations~\ref{enrgyconv2ndcase} and~\ref{energyconvcond2}, we can find the following relation for $f$, i.e.,
\beq
\label{f-relation}
f = \frac{\rho_{+}(T_c) - \rho_{+}(T_n)}{L},
\eeq
where $L~(= \rho_{+}(T_c) - \rho_{-}(T_c))$ denotes the transition latent heat at $T= T_c$.
Finally, the process of the third stage, i.e., the phase coexistence stage, occurs very quickly such that the total entropy again remains conserved which leads to the following relation:
\beq
\label{entrpyconvcase2}
a_s^3[(1-f)s_{+}(T_c) + fs_{-}(T_c)] = a_f^3 s_{-}(T_c).
\eeq
The left- and right-hand sides of the above relation correspond to the total entropy at the beginning and at the end of this stage, respectively. It can be expressed as
\beq
\label{entrpyconvcase3}
\bigg(\frac{a_s}{a_f}\bigg)^3 = \frac{1 - \Delta s/s_{+}(T_c)}{1 - f\Delta s/s_{+}(T_c)},
\eeq
where $\Delta s = s_{+}(T_c) - s_{-}(T_c)$. Therefore, by combining equations~\ref{dilutioncase1} and~\ref{entrpyconvcase3}, it is possible to find the dilution factor for this scenario and it is given by
\beq
\label{dilutioncase2}
d \equiv \bigg(\frac{a_f}{a_i}\bigg)^3 = \bigg(\frac{1 - f\Delta s/s_{+}(T_c)}{1 - \Delta s/s_{+}(T_c)}\bigg)\bigg(\frac{s_{+}(T_c)}{s_{+}(T_n)} \bigg).
\eeq
In this work, we use this relation to estimate the dilution of the DM relic density caused by an FOPT in section~\ref{result}.

In addition, FOPT in the early Universe could generate stochastic GW. Such a GW signal can be estimated from the four portal parameters that are related to the phase transition in the scalar potential and those are $T_n, \alpha, \beta/H_n$ and $v_w$ where 
`$\alpha$' is a dimensionless quantity which is related to the energy budget of the FOPT,
$\beta/H_n$ estimates the inverse duration of the phase transition and $v_w$ is the bubble wall velocity. Definitions of these parameters are shown in Appendix~\ref{GWexpression}. 
Such a GW
from an FOPT is generally produced via three distinct processes: bubble wall collisions~\cite{Kosowsky:1992rz, Kosowsky:1991ua}, long-standing
sound waves in the plasma~\cite{Hindmarsh:2013xza, Giblin:2013kea, Giblin:2014qia, Hindmarsh:2015qta} and magnetohydrodynamic (MHD) turbulence~\cite{Caprini:2006jb, Kahniashvili:2008pf, Kahniashvili:2008pe, Kahniashvili:2009mf, Caprini:2009yp, Kisslinger:2015hua}. 
Among all these, the sound wave contribution is the dominant contribution to the peak of the GW intensity. Various relations from that computation are presented in Appendix~\ref{GWexpression}.
We use these relations to study the connection between the dilution of the DM relic density and the produced stochastic GW resulting from an FOPT. 

Before ending this section, note that there are some prerequisites in estimating the dilution factor. 
It is related to the assumption that the released latent heat after the supercooling reheats the Universe which brings the temperature back to a temperature just close to $T_c$. Observe that the value of the variable $f$, that estimates the plasma volume fraction of the low-temperature phase at the starting of the phase coexistence stage, cannot be larger than 1 in equations~\ref{enrgyconv2ndcase} and~\ref{energyconvcond2}. However, in some situations, particularly when the dilution factor is large, the value of $f$ can exceed the value 1 using the relation in equation~\ref{f-relation}. The underlying assumption made above has the explanation behind this. When $T_n << T_c$, the large amount of released latent heat cannot bring the system to a temperature close to $T_c$. In this scenario, for the correct estimation of the dilution factor, $T_n$ needs to be replaced by the reheating temperature $T_r$ in equation~\ref{dilutioncase2}. Although $T_r$ estimation is beyond the scope of this work as such a calculation  
consider specific kinetic processes~\cite{Megevand:2007sv}. If we neglect the contributions from the terms that includes $f$, i.e., considering $f= 1$, in equation~\ref{dilutioncase2} the relation of dilution factor becomes $
d \equiv \bigg(\frac{a_f}{a_i}\bigg)^3 = \bigg(\frac{s_{+}(T_c)}{s_{+}(T_n)} \bigg) = \bigg(\frac{T_c}{T_n}\bigg)^3 = \bigg(1+\frac{\Delta T_{\text{cn}}}{T_n}\bigg)^3
$~\cite{Xiao:2022oaq}.
This relation actually estimates a much-enhanced dilution factor. In addition, in the realistic scenario, friction will drastically reduce the mobility of bubble walls in the plasma-wall system. The collisions of the walls of the bubbles of the truly broken phase also release entropy and enhance the dilution factor~\cite{Wainwright:2009mq}. 
Therefore, refuting the underline assumption does not imply a reduction in the dilution factor that we estimate in this work. 
\section{The theoretical scenario}
\label{model}
The current model is an extension of the SM with a $SU(2)_L$ scalar triplet, $\Delta$, with hyper-charge, $Y=2$, and a complex scalar singlet, $S$, which transforms non-trivially under a $Z_3$-discrete symmetry~\cite{Ghosh:2022fzp, Yang:2021buu}.
The discrete $Z_3$- symmetry stabilises the singlet field, i.e., it does not develop $\vev$ on EWSB and becomes the DM candidate of the model.
In Table~\ref{modeltable}, we present various quantum numbers of various scalars of the scenario
under the extended gauge group $SU(3)_C\times SU(2)_L \times U(1)_Y \times Z_3$.
The Lagrangian of the model is given by
\begin{eqnarray}
{\cal{L}}_{\mathrm{tot}}={\cal{L}}_{\mathrm{Kinetic}} + {\cal{L}}_{\mathrm{Yukawa}} - {\cal{V}}(H,\Delta,S).
\end{eqnarray}
where
\begin{eqnarray}\label{kinetic}
{\cal{L}}_{\mathrm{Kinetic}} &=& (D^{\mu}H)^{\dagger} D_{\mu}H +(D^{\mu}\Delta)^{\dagger} D_{\mu}\Delta + (\partial^\mu{S})^{\dagger}\partial_{\mu}S\ , \\
{\cal{L}}_{\mathrm{Yukawa}} &=& {\cal{L}}^{\mathrm{SM}}_{\mathrm{Yukawa}} - (y_L)_{\alpha\beta} ~\overline{L_\alpha^c}~i \sigma^2 \Delta ~L_\beta + h.c.\ ,
\end{eqnarray}
with
\bea
D_{\mu}H &=& \Big(\partial_\mu - i g_2 \frac{\sigma^a}{2} W_{\mu}^a-i g_1 \frac{Y_{H}}{2} B_{\mu}\Big)H , \nonumber \\
D_{\mu}\Delta &=& \partial_{\mu}\Delta - i g_2 \left[\frac{\sigma^{a}}{2} W_{\mu}^a ,\Delta\right] - {i g_1} \frac{Y_\Delta}{2} B_{\mu}\Delta ~.
\eea
Here, $y_L$ is a $3 \times 3$ complex symmetric matrix. $g_1$ and $g_2$ are the 
$U(1)$ and the $SU(2)$ gauge couplings, respectively. The heavy triplet scalar helps us to
generate neutrino masses through the standard type-II seesaw mechanism. Neutrino masses are generated once the neutral triplet scalar gets  non-zero $\vev$ after the EWSB.
The representations of the Higgs doublet ($H$), the triplet scalar ($\Delta$) and the singlet scalar ($S$) are shown in table~\ref{modeltable}.
After the EWSB,  $\vev$  of the neutral $CP$-even field of $H$ ($\Delta$) is denoted by $v_d$ ($v_t$) where  $v=\sqrt{v_d^2+ 2 v_t^2} =246$~GeV.
\begin{table}[t]
\resizebox{\linewidth}{!}{
 \begin{tabular}{|c|c|c|c|}
\hline \multicolumn{2}{|c}{Fields}&  \multicolumn{1}{|c|}{ $\underbrace{ SU(3)_C \otimes SU(2)_L \otimes U(1)_Y}$ $\otimes  Z_3  $} \\ \hline
Complex Scalar DM & $s=\frac{1}{\sqrt{2}}(h_s + i a_s)$ & ~~1 ~~~~~~~~~~~1~~~~~~~~~~~~0~~~~~~~~~$e^{i\frac{2\pi}{3}}$ \\
\hline
\hline
Scalar Triplet & $\Delta=\left(\begin{matrix} \frac{\Delta^+}{\sqrt{2}} & \Delta^{++} \\ \frac{1}{\sqrt{2}}\big(h_t + i a_t \big) & -\frac{\Delta^+}{\sqrt{2}} \end{matrix}\right)$ & ~~1 ~~~~~~~~~~~3~~~~~~~~~~2~~~~~~~~~~1 \\
\hline
Higgs doublet & $H=\left(\begin{matrix} G^+ \\ \frac{1}{\sqrt{2}}\big(h_d + i a_d \big) \end{matrix}\right)$ & ~~1 ~~~~~~~~~~~2~~~~~~~~~~1~~~~~~~~~~1 \\
\hline
\end{tabular}
}
\caption{
Assignments of charges on the scalar fields under the gauge group $\mathcal{G} \equiv \mathcal{G}_{\rm SM} \otimes Z_3$  where $\mathcal{G}_{\rm SM}\equiv SU(3)_C \otimes SU(2)_L \otimes U(1)_Y$. The hypercharges ($Y$) are obtained by using the relation: $Q= I_3 +\frac{Y}{2}$, where  $Q$ is the electromagnetic charge and $I_3$ is the third component of isospin. }
    \label{modeltable}
\end{table}
The scalar potential of the model is given by:
\bea
V(H,\Delta, S)&=& -\mu_{H}^2 (H^\dagger H)+ \lambda_{H} (H^\dagger H)^2 ~ \nonumber \\
&& + \mu_{\Delta}^2 {\rm Tr}\left[\Delta^{\dagger}\Delta\right]  + \lambda_{1} \left(H^{\dagger}H\right){\rm Tr}\left[\Delta^{\dagger}\Delta\right] + \lambda_2 \left({\rm Tr}[\Delta^{\dagger}\Delta]\right)^2 \nonumber \\
& & +\lambda_3 ~{\rm Tr} [\left(\Delta^{\dagger}\Delta\right)^2] + \lambda_4 ~\left(H^{\dagger}\Delta\Delta^{\dagger}H \right)+ \left[\mu\left(H^T i \sigma^2 \Delta^{\dagger}H\right)+h.c.\right] + \mu_{S}^2 ~ |S|^2 ~ \nonumber \\
& & + \lambda_{S} |S|^4 + \frac{\mu_3}{3!} \Big(S^3 + {S^*}^3 \Big)
+ \lambda_{S H} H^\dagger H  (S^* S) + \lambda_{S \Delta}~ {\rm Tr}\left[\Delta^{\dagger}\Delta\right](S^* S)~~.
\label{scalpot}
\eea
In this work, we consider all the Lagrangian parameters to be real. Thus, there is no spontaneous $CP$ violation in our scenario. The bare masses ($\mu_H^2$ and $\mu_\Delta^2$) can be obtained by minimizing the scalar potential for the field values at the vacuum.
The mass of the singlet scalar DM after the EWSB is given by
\bea
m_S^2= \mu_S^2 + \frac{1}{2} \lambda_{S H} v_d^2 + \frac{1}{2} \lambda_{S \Delta} v_t^2  ~.
\eea
On EWSB, the doublet and the triplet scalars would mix. The neutral scalar mass-matrix, on diagonalization, gives rise to
two $CP-$even scalars $h^0$ and $H^0$. These are given by,
\bea
h^0 = h_d ~\cos\theta_t + h_t ~\sin\theta_t  ~, ~~~~
H^0 = -h_d~ \sin\theta_t + h_t ~\cos\theta_t \, ,
\eea
where $\theta_t$ represents the mixing angle of the diagonalizing matrix and is given by
\bea
\label{thetat}
\tan2\theta_t &=& \frac{\sqrt{2}\mu v_d -(\lambda_1+\lambda_4)v_d ~v_t}{M_\Delta^2-\frac{1}{4}\lambda_H v_d^2 + (\lambda_2+\lambda_3) v_t^2} \, \, ,
\label{tanthetat}
\eea
where $M_{\Delta}^2=\frac{\mu  v_d^2}{\sqrt{2} v_t}$.
The corresponding mass eigenvalues of these physical eigenstates are given by
\bea
m_{h^0}^2&=& \left(M_\Delta^2+2 v_t^2 (\lambda_2+\lambda_3)\right) \sin ^2\theta_t + 2 \lambda_H v_d^2 \cos ^2\theta_t-\frac{v_t \sin2\theta_t \left(2 M_\Delta^2 - v_d^2 (\lambda_1+\lambda_4)\right)}{v_d} \, ,   \nonumber \\ 
m_{H^0}^2&=& \left(M_\Delta^2+2 v_t^2 (\lambda_2+\lambda_3)\right) \cos ^2\theta_t + 2 \lambda_H v_d^2 \sin^2\theta_t+\frac{v_t \sin2\theta_t \left(2 M_\Delta^2 - v_d^2 (\lambda_1+\lambda_4)\right)}{v_d} .\nonumber \\
\label{MCPE}
\eea
For $m_{h^0} < m_{H^0}$, the lighter eigenstate $h^0$ mimics the SM Higgs boson with mass $\mhsm=125$ GeV.
Similarly, the mixing between two $CP$-odd states ($a_d$, $a_t$) would project out the massless Nambu-Goldstone mode and one massive pseudoscalar, $A^0$ with its mass given by
\bea
m_{A^0}^2&=& \frac{M_\Delta^2 \left(4 v_t^2+v_d^2\right)}{v_d^2}~~ .
\label{MCPO}
\eea
A similar orthogonal rotation in the singly-charged scalar sector comprising of $G^\pm$ and $\Delta^\pm$ fields would lead to one massless Nambu-Goldstone state which shows up as the longitudinal components of the massive $W^\pm$-bosons and the charged scalar $H^\pm$. The mass of $H^\pm$ is given by
\bea
m_{H^\pm}^2 &=& \frac{\left(2 v_t^2+v_d^2\right) \left(4 M_\Delta^2 - \lambda_4 v_d^2\right)}{4 v_d^2}  .
\label{MSC}
\eea
The mass of the doubly charged scalar $ H^{\pm\pm} (\equiv \Delta^{\pm\pm} )$ is given by
\bea
m_{H^{\pm\pm}}^2 &=& M_\Delta^2-\lambda_3 v_t^2-\frac{\lambda_4 v_d^2}{2}   ~~.
\label{MDC}
\eea
Note that the constraint on $m_{H^{\pm\pm}}$ from the various heavy Higgs searches at the LHC mostly depends on $v_t$, $m_{H^{\pm\pm}}$ and the mass difference $\Delta m$ $(= m_{H^{\pm\pm}} - m_{H^{\pm}}$). At the limit $v_t^2/v_d^2 <<1$, $\Delta m$ can be found from the following relation:
\bea
\label{trplmassdiff}
\big(m_{H^{\pm\pm}}^2-m_{H^\pm}^2 \big) \approx -\frac{\lambda_4 v_d^2}{4}~; ~~ \big( m_{H^\pm}^2-m_{A^0}^2 \big) \approx -\frac{\lambda_4 v_d^2}{4} ~;~~
m_H  \approx m_A^0 \approx M_\Delta .
\eea
Thus, sign of $\lambda_4$ generates two types of the mass hierarchy among $H^{\pm\pm}$, $H^{\pm}$, $H^{0}/A^0$. For $\lambda_4$ $<0$ ($>0$) the mass ordering is $m_{H^{\pm\pm}} > m_{H^\pm} > m_{H^0,A^0}$ ($m_{H^{\pm\pm}} < m_{H^\pm} < m_{H,A^0}$).

Various theoretical constraints such as stability, unitarity, perturbativity and the latest experimental constraints from the dark sector  as well as from the various LHC searches, the constraints from the flavor and neutrino sectors and the constraints from the electroweak precision data observables such as $\rho$, $S$, $T$, $U$ parameters have been discussed in detail in reference~\cite{Ghosh:2022fzp}. 

In the present work, we focus on that kind of parameter space region exhibiting an FOEWPT in the early Universe. Such a scenario prefers the triplet sector scalars to be lighter. 
However, the null results from various collider searches that have been carried out over the years at the LHC~\cite{ATLAS:2012hi,Chatrchyan:2012ya,ATLAS:2014kca,Khachatryan:2014sta,CMS:2016cpz,CMS:2017pet,Aaboud:2017qph,CMS:2017fhs,Aaboud:2018qcu,Aad:2021lzu} to hunt those triplet-like scalar excitations, strongly constrain the parameter space of the type-II seesaw sector of the present model. Such a bound on $m_{H^{\pm\pm}}$ mostly depend on $\Delta m$, $v_t$ and the value of $m_{H^{\pm\pm}}$.
In the scenario when $\Delta m = 0$, the ATLAS collaboration finds the most stringent bound to be $m_{H^{\pm\pm}} \gtrsim 870$ GeV. This was accomplished by making the assumption that BR($H^{\pm\pm} \rightarrow \mu^{\pm} \mu^{\pm}$) is 100$\%$~\cite{Aaboud:2017qph}, which corresponds to a smaller $v_t$ region, using 36.1 fb$^{-1}$ data at $\sqrt{s}=13$ TeV. On the other hand, a different ATLAS analysis assuming  100$\%$ branching fraction of $H^{\pm\pm}$ to the $W^{\pm} W^{\pm}$ final state, that corresponds to the larger $v_t$ region, excludes on  $m_{H^{\pm\pm}} \gtrsim 350$ GeV~\cite{Aad:2021lzu} with 139 fb$^{-1}$ data.
In the case of $\Delta m < 0$ the bound is much stronger. In this scenario, the effective production cross-section of $H^{\pm\pm}$ at the LHC increases as it can produce from the cascade decays of $H^0/A^0$ and $H^\pm$. 
By recasting several ATLAS and CMS results, it has been shown in a recent study~\cite{Ashanujjaman:2021txz} that the existing exclusion limit for $m_{H^{\pm\pm}}$ is 1115 (1076) GeV for $\Delta m = -10$ ($-30$) GeV with $v_t \sim 10^{-5} - 10^{-6}$ GeV. 
On the contrary, the bound on $m_{H^{\pm\pm}}$ relaxes a lot in the case of $\Delta m > 0$.

The scenario with $\Delta m>$~0, the neutral triplet-like scalar $H^0$ and $A^0$ mostly decay to invisible neutrinos for $v_t <$ $10^{-4}$ GeV~\cite{Ashanujjaman:2021txz}. Thus the produced $H^{\pm\pm} H^{\mp\mp}$ and $H^{++}H^{--}$ yields soft-leptons/jets in the final states. Since the particles are very soft, it is difficult to reconstruct them at the LHC. Recently, by recasting various ATLAS and CMS analyses in reference~\cite{Ashanujjaman:2021txz}, it has been shown that LHC would be unable to constrain the parameter space of $\Delta m =$ 10 GeV (30 GeV),  $v_t$ around $10^{-3}$ GeV to $10^{-5}$ GeV ($10^{-3}$ GeV to $10^{-6}$ GeV) and $m_{H^{\pm\pm}} \gtrsim$ 200 GeV at 3 ab$^{-1}$ integrated luminosity data.

Therefore, the region of parameter space of the triplet sector that we consider for this work corresponds to,~\footnote{In this article, we focus on a similar region of parameter space of the type-II seesaw model that is considered in reference~\cite{Ghosh:2022fzp}. Note that, from the EW
precision data $\Delta m$ is upper bounded and it has to be below 40 GeV.}
\bea
\{10\, \text{GeV} <\Delta m< 40\, \text{GeV}\} \Rightarrow \lambda_4 <0~;\, \{10^{-6} \, \text{GeV} < v_t < 10^{-3}\ \, \text{GeV}\}; \, \{m_{H^{\pm\pm}} > 200 \, \text{GeV}\}.~~\nonumber \\
\label{MDC}
\eea
The precision study of $\hsm$ at the LHC restricts the large mixing of the doublet-like scalar with the triplet-like neutral scalar as the observed $\sim$ 125 GeV Higgs boson is more SM-like. Therefore, we limit our scan to tiny $\sin\theta_t$ values to allow minimum mixing and maintain the nature of the Higgs boson SM-like. However, loop-induced decays of the SM-like Higgs boson ($\hsm$), such as $\hsm \rightarrow \gamma \gamma$, can still deviate significantly from the experimentally observed limits of the signal strength of the same decay channel even at small $\sin\theta_t$  value. The variation of the signal strength of the $\hsm \rightarrow \gamma \gamma$ decay mode in the above-discussed region of parameter space in equation~\ref{MDC} is discussed in reference~\cite{Ghosh:2022fzp, Zhou:2022mlz}.
Note that, at relatively small triplet-like charged Higgs bosons and larger portal couplings ($\lambda_1, \lambda_4$), the decay width of the SM Higgs into a photon pair significantly deviates from the observed limits~\cite{ATLAS:2022tnm, CMS:2021kom}.

 \begin{figure}[htb!]
   \begin{center}
    \begin{tikzpicture}[line width=0.6 pt, scale=2.1]
\draw[dashed] (-1.8,1.0)--(-0.8,0.5);
\draw[solid] (-1.8,-1.0)--(-0.8,-0.5);
\draw[dashed] (-0.8,0.5)--(-0.8,-0.5);
\draw[dashed] (-0.8,0.5)--(0.2,1.0);
\draw[solid] (-0.8,-0.5)--(0.2,-1.0);
\node at (-2.1,1.1) {$S$};
\node at (-2.1,-1.1) {$n$};
\node at (-0.5,0.07) {$h^0,H^0$};
\node at (0.5,1.2) {$S$};
\node at (0.5,-1.2) {$n$};
     \end{tikzpicture}
 \end{center}
\caption{Feynman diagram illustrating the spin-independent DM-nucleon scattering process mediated by the $CP$-even scalars, $h^0$ and $H^0$, for the scalar DM.}
\label{DD}
 \end{figure}
Numerous experimental efforts have been made to detect DM in laboratory experiments. Among these, DMDD experiments search for signals arising from the scattering of incoming DM particles with nuclei in target materials, leading to nuclear recoils. The spin-independent DM-nucleon scattering cross-section ($\sigma_{\rm DD}^{\rm SI}$), mediated via $t$-channel exchanges of CP-even scalars ($h^0$ and $H^0$), as shown in Fig.~\ref{DD}, is given in the zero-momentum transfer limit ($q \to 0$) as:
\begin{equation}
\sigma_{\rm DD}^{\rm SI} = \Big( \frac{\Omega_S h^2}{\Omega_{\rm DM} h^2} \Big) \frac{1}{4\pi} \bigg( \frac{f_n \mu_n}{m_S} \bigg)^2 \bigg( \frac{m_n}{v_d} \bigg)^2 \bigg[ \frac{\lambda_{D1}}{m_{h^0}^2} + \frac{\lambda_{D2}}{m_{H^0}^2} \bigg]^2,
\label{eq:sigmadd}
\end{equation}
where $\xi_{\rm DM} = \Omega_S h^2 / \Omega_{\rm DM} h^2$ represents the fractional DM density. The couplings $\lambda_{D1}$ and $\lambda_{D2}$ are defined as:
\begin{align}
\lambda_{D1} &= \big( \lambda_{SH} v_d \cos\theta_t + \lambda_{S\Delta} v_t \sin\theta_t \big) \cos\theta_t \xrightarrow{\theta_t \to 0} \lambda_{SH} v_d, \\
\lambda_{D2} &= \big( -\lambda_{SH} v_d \sin\theta_t + \lambda_{S\Delta} v_t \cos\theta_t \big) \sin\theta_t \xrightarrow{\theta_t \to 0} 0.
\end{align}

Here, $\mu_n = \frac{m_S m_n}{m_S + m_n}$ denotes the reduced mass of the DM-nucleon system, where $m_n = 0.946~\mathrm{GeV}$ is the nucleon mass, and $f_n = 0.28$ is the nucleon form factor~\cite{Alarcon:2012nr}. In the limit $\theta_t \to 0$, $\sigma_{\rm DD}^{\rm SI}$ primarily depends on the Higgs portal coupling $\lambda_{SH}$ and the DM mass $m_S$, i.e., $ \sigma_{\rm DD}^{\rm SI} \propto \frac{\lambda_{SH}^2}{m_S^4}$.
Latest experimental results from XENON-1T~\cite{Aprile:2018dbl}, PandaX-4T~\cite{PandaX-4T:2021bab}, and LUX-ZEPLIN (LZ)~\cite{LZCollaboration:2024lux} experiments have imposed stringent constraint on DMDD scattering cross-section. These bounds can be further expressed in terms of the parameter space of the model.
For a given DM mass, these constraints restrict $\lambda_{SH}$.
Note that, for sufficiently small $\lambda_{SH}$, the DMDD cross-section can fall below the neutrino floor, as shown in Fig.~5 of Ref.~\cite{Ghosh:2022fzp}.

At the same time, DM indirect detection (DMID) experiments, such as the space-based Fermi Large Area Telescope (Fermi-LAT)~\cite{Fermi-LAT:2017bpc} and the ground-based MAGIC~\cite{MAGIC:2016xys} telescope, search for gamma-ray signals originating from DM interactions. These gamma rays are generated through the production of the SM particles resulting from DM annihilation or decay in the local universe. For WIMP-like DM, the emitted photons lie within the gamma-ray regime, serving as ideal messengers for indirect detection. 

The total gamma-ray flux resulting from DM annihilation into SM particle pairs ($X \overline{X}$) within a specific energy range is expressed as~\cite{MAGIC:2016xys}:
\begin{eqnarray}
\label{eq:id}
\Phi_{X \overline{X}} &=& \frac{1}{4\pi} \frac{\langle \sigma v \rangle_{SS \to X \overline{X}}}{2 m_{S}^2} \int_{E_{\rm min}}^{E_{\rm max}} \frac{dN_\gamma}{dE_\gamma} dE_{\gamma} \underbrace{\int dx ~\rho_S^2\big(r(b,l,x)\big)}_{J} ~,
\end{eqnarray}
where $X := \{\mu^-,\tau^-,b,W^-\}$. The $J$ factor encodes the astrophysical information about DM density distribution, while $\frac{dN_\gamma}{dE_\gamma}$ represents the photon energy spectrum. Using these flux measurements and astrophysical inputs, experimental limits are placed on the thermal annihilation cross-section of DM into SM channels, such as $\mu^+\mu^-,~\tau^+\tau^-,~b\overline{b},~W^+W^-$.
To compare theoretical predictions with experimental limits, the flux can be reformulated as:
\begin{eqnarray}
\label{indirectcross}
\langle \sigma v \rangle_{SS \to X\overline{X}}^{\rm eff} &=& \big(\Omega_S/\Omega_{\rm DM} \big)^2 \langle \sigma v \rangle_{SS \to X\overline{X}} = \xi_{\rm DM}^2 \langle \sigma v \rangle_{SS \to X\overline{X}}~.
\end{eqnarray}
Here, $\xi_{\rm DM}^2$ scales the cross-section by the fractional DM density. In this framework, the annihilation cross-section $\langle \sigma v \rangle_{SS \to X\overline{X}}$ is predominantly mediated by $s$-channel SM-like Higgs $h^0$ exchange, scaling as $\propto \lambda_{SH}^2 / m_S^2$ for $\sin\theta_t \to 0$. Null results from indirect searches impose stringent constraints on the annihilation cross-sections, which can be translated into restrictions on the model parameters. 
Later, we will discuss the impact of the dilution factor on DMDD and DMID experimental constraints, along with its correlation to the model parameter space, in Section~\ref{dilutionfactor}.

In this article, we analyze the thermal history of the scalar potential and determine the region of the parameter space that features an FOEWPT in the early Universe. After such an FOEWPT, the broken phase eventually evolves to the EW  minimum of the potential at $T = 0$. 
We use $\mathbf{CosmoTransitions}$~\cite{Wainwright:2011kj} to study the evolution of various minima with temperature as well as the transition between the various minima of the scalar potential at finite temperature.
The theoretical details for the study of phase transition in the early Universe, such as the field-dependent mass relations of various degrees of freedom and the daisy coefficients (to estimate the thermal mass corrections) are described in detail in reference~~\cite{Ghosh:2022fzp}. Furthermore, we analyse the dilution factor of the thermal relic density caused by such an FOEWPT.
 Various DM observables, such as its relic density, its freeze-out temperature and its spin-independent direct detection cross-section have been estimated using \micromegas~{\tt (v4.3)}~\cite{Belanger:2006is}.
\section{Result}
\label{result}
With the description of the model and various experimental constraints in section~\ref{model}, we move on to present our results.
we discuss the feasibility of a scalar DM with a mass around/above~1~TeV in our scenario, as well as the range of the freeze-out temperatures  for such a heavy DM in subsection~\ref{TevscaleDM}. 
Furthermore, in subsection~\ref{dilutionfactor}, we estimate the dilution factor of the DM relic density due to an FOEWPT. We discuss the region of parameter space of this model where the dilution factor becomes significantly large.
As pointed out in the introduction, we further study the dependencies of the dilution factor on the model parameters, the nucleation temperature, the strength
and the duration of the phase transition. 
Finally, in subsection~\ref{GWdilutioncorrelation}, we discuss the connection between the production of stochastic GW and the dilution factor as a result of an FOEWPT.
\subsection{TeV-scale scalar DM and the freeze-out temperature}
\label{TevscaleDM}
\begin{figure}[t!]
\begin{center}
\hskip-18pt
\includegraphics[height=5.7cm,width=0.51\linewidth]{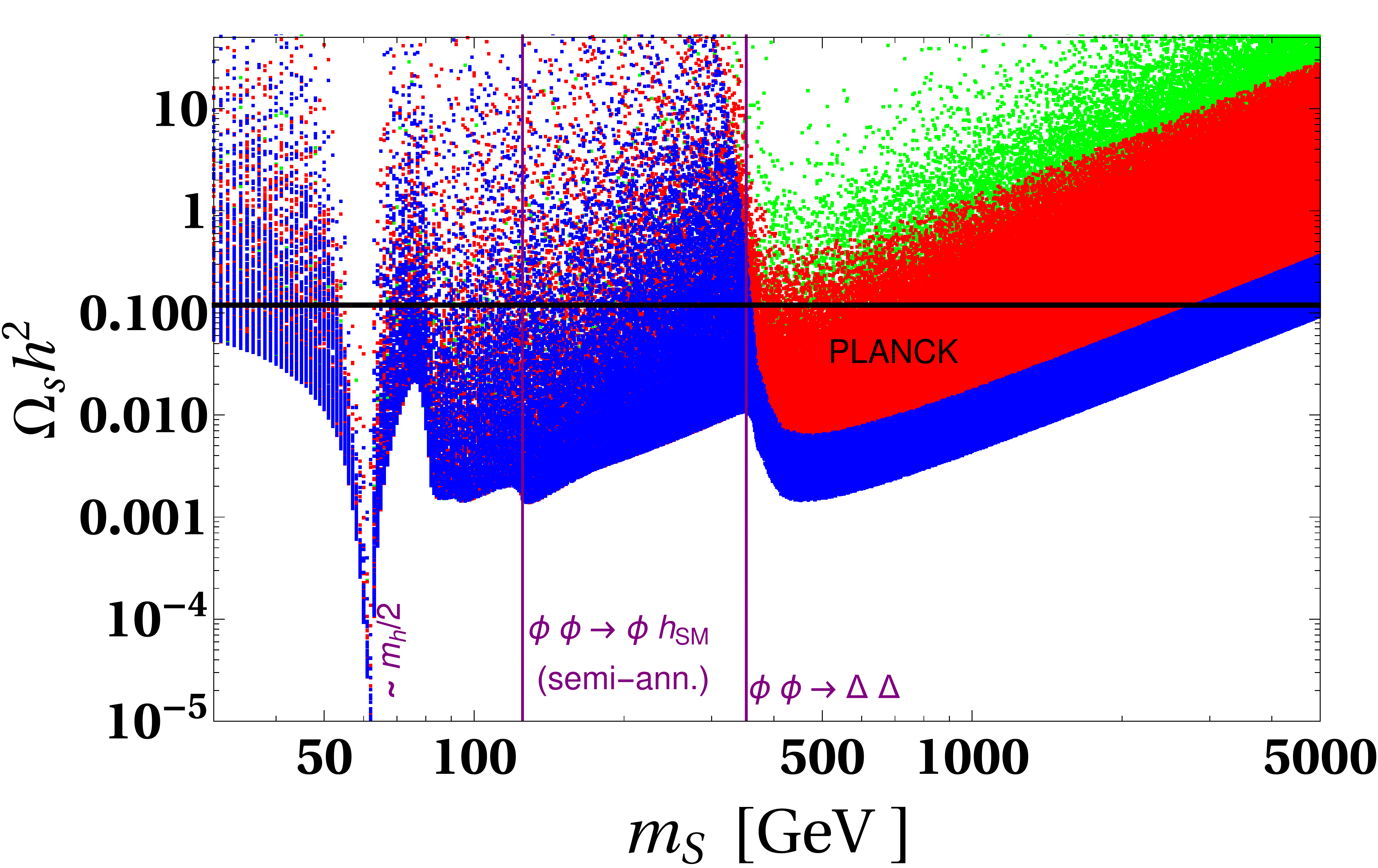}
\hskip 2pt
\includegraphics[height=5.8cm,width=0.51\linewidth]{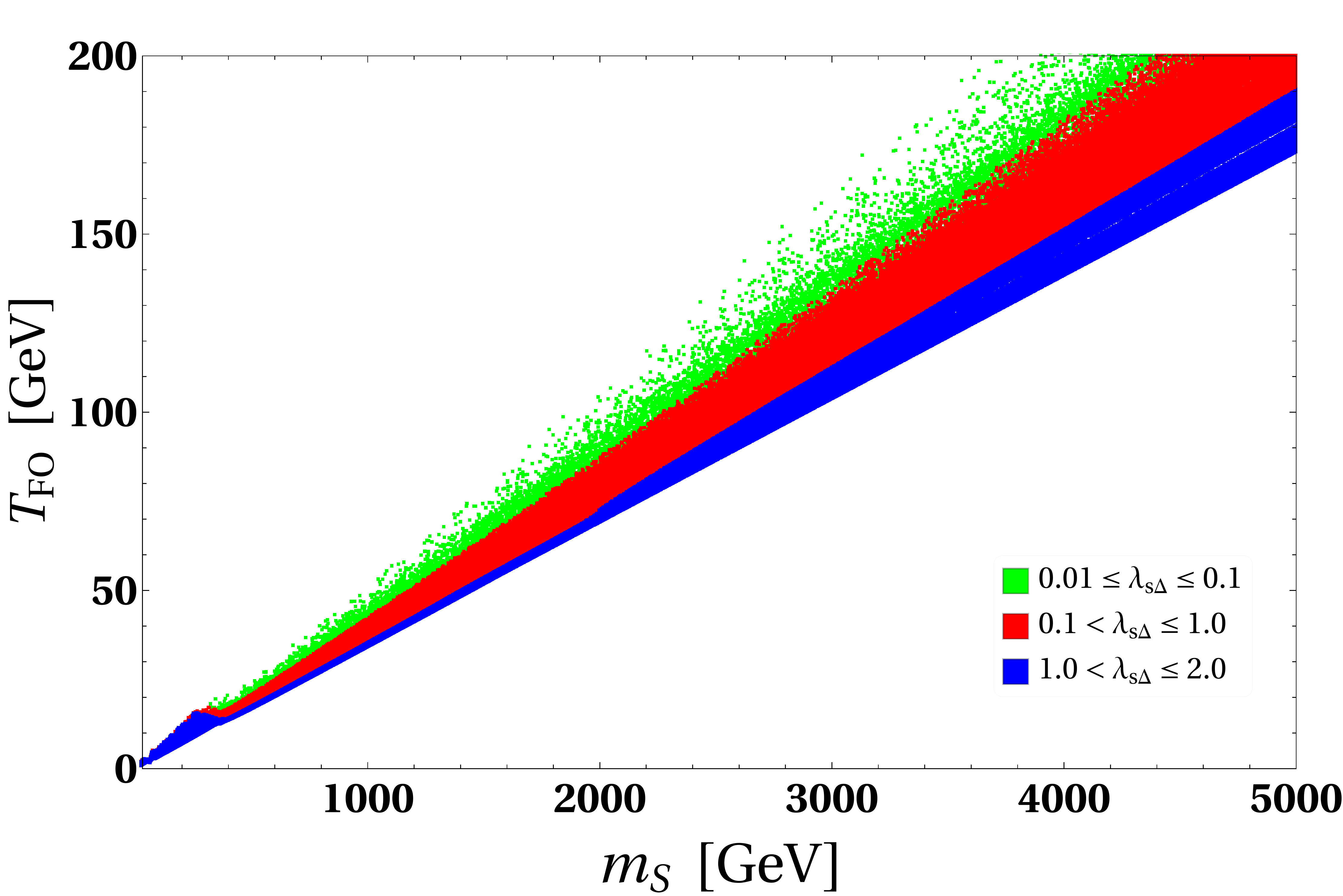}
\caption{[Left plot:] variation of relic density ($\Omega_s h^2$) as a function of DM mass ($m_s$) in GeV for different ranges $\lambda_{S \Delta}$ : $0.01 \leq \lambda_{S \Delta} \leq 0.1$ (green), $0.1 < \lambda_{S \Delta} \leq 1.0$ (red), $1.0 < \lambda_{S \Delta} \leq 2.0$ (blue) and $0.001 < \lambda_{SH} \leq 0.5$. [Right plot:] variation of freeze-out temperature ($T_{\text{FO}}$) with $m_s$ for same range of $\lambda_{S\Delta}$ and $\lambda_{SH}$. In these two plots $\mu_3 = m_s$ and the triplet sector parameters are fixed at $m_{H^0,A^0} = 368$~GeV , $m_{H^\pm} = 388$~GeV, $m_{H^{\pm\pm}} = 406$~GeV, $\sin\theta_t=-1.0 \times 10^{-6}$. In the left plot, on
the vertical axis $\Omega_s h^2$ is truncated at the value of 50 and on the horizontal axis $m_s$ is considered up to 5 TeV. Black horizontal lines indicate the Planck observed DM relic density, i.e.,  $\Omega_{\text{DM}} h^2=0.120 \pm 0.001$. The vertical black lines indicate the opening of various DM annihilation channels. In the right plot, the horizontal axis is truncated at the value of $T_{\text{FO}}= 200$ GeV.}
\label{fig:relic-annihilation}
\end{center}
\vspace{-0.5cm}
\end{figure}
The phenomenology of the scalar DM in our situation is discussed in this subsection.  The DM annihilates in three ways to freeze out in the early Universe and those processes are given by: DM ~annihilation$ \hspace{0.3cm} (a)~ S~ S^{*} ~\to {\rm SM~~SM}$, (b) $S~ S^{*} ~\to {\rm X~~Y}; ~~~\{\rm X,Y\}=\{H^0,A^0,H^\pm, H^{\pm\pm}\}$ and DM ~semi-annihilation (c) $~ S~ S ~\to  S~U; ~~\{U\}=\{\hsm,H^0\} $.
The DM Higgs portal interactions $\lambda_{S H}(S^* S)(H^\dagger H)$ and $\lambda_{S \Delta}(S^* S){\rm Tr}[\Delta^\dagger \Delta]$ control the DM-annihilation processes (a) and (b), respectively, whereas the DM semi-annihilation processes are initiated via the cubic interaction term $\mu_3(S^3 + {S^*}^3)$ in the scalar potential.\footnote{Note that at small $v_t$-limit the DM semi-annihilation process $S~ S \to  S~H^0$ and the $S$-channel neutral triplet-like scalar $H^0$-mediated DM-annihilation processes are highly suppressed.} Thus, the DM relic density varies as follows:
 \bea
 \label{relic-annihilation}
 \Omega_{S}h^2 \propto \frac{1}{\langle \sigma v \rangle _{S S \to {\rm SM~SM}}+\langle \sigma v \rangle _{S S \to S U}+\langle \sigma v \rangle _{S S \to X~Y}}~.
 \eea
In the left plot of figure~\ref{fig:relic-annihilation}, we illustrate the variation of $\Omega_S h^2$ as a function of $m_S$ while fixing the triplet-sector parameters at
$m_{H^0,A^0} = 368$~GeV, $m_{H^\pm} = 388$~GeV, $m_{H^{\pm\pm}} = 406$~GeV and $\sin\theta_t=-1.0 \times 10^{-6}$ while the DM-Higgs interaction couplings are varied in the range of $0.001 < \lambda_{SH} < 0.5$, $0.01 <\lambda_{S\Delta} < 2$ and  $0 < m_s< 5$~TeV.
Such a choice of parameters is considered to understand the dependencies of the phenomenology of DM on the model parameters.
In the right plot of figure~\ref{fig:relic-annihilation}, we present the variation of the freeze-out temperature ($T_{\text{FO}}$) with $m_S$.
Different colors show different ranges of $\lambda_{S\Delta}$ in these two plots. Colors in green, red and blue indicate $0.01 \leq \lambda_{S \Delta} \leq 0.1$, $0.1 < \lambda_{S \Delta} \leq 1.0$ and $1.0 < \lambda_{S \Delta} \leq 2.0$, respectively.
The black dotted horizontal lines indicate the measured density of DM relics from the PLANCK experiment, i.e., $\Omega_{\text{DM}} h^2 = 0.120 \pm 0.001$~\cite{Planck:2018vyg}.
The equation~\ref{fig:relic-annihilation} suggests that the relic density might rapidly decrease with growing DM mass as a result of the opening of new number-changing processes at a certain threshold DM mass. 
Such a phenomenon shows up in the left plot. 

In the scenario where the DM mass is larger than the triplet-like scalar masses, the DM annihilation processes mentioned in the category `b' open up and the four-point contact interaction term, $\lambda_{S \Delta}(S^* S){\rm Tr}[\Delta^\dagger \Delta]$, starts to contribute.
A larger $\lambda_{S\Delta}$ enhances the DM annihilation cross-sections and beyond a certain value, the DM annihilation cross-section of the processes of the category `b' becomes the dominant one compared with the other annihilation processes.
Note that in the limit of small $v_t$ and $\sin\theta_t$, the contribution of the $H^0$-mediated $t$-channel process to the DM direct detection spin-independent (DMDD-SI) cross-section is negligible.
Therefore, the DMDD-SI cross-section mostly depends on $\lambda_{SH}$ (mediated via $\hsm$) and has an insignificant effect on the variation of $\lambda_{S\Delta}$.
Thus, for a heavier DM, the region of parameter space with larger $\lambda_{S\Delta}$ can meet the experimentally observed DM relic density criteria without increasing the DMDD-SI cross-section significantly. 
As a result, this model can accommodate a heavy ($\sim$TeV-scale) DM complying with various DM experimental constraints~\cite{LZCollaboration:2024lux,MAGIC:2016xys}.
The variation of  $T_{\text{FO}}$ with $m_S$ in the right plot of figure~\ref{fig:relic-annihilation} indicates that heavy $m_S$ corresponds to larger $T_{\text{FO}}$. 
It follows roughly $T_{\text{FO}} \sim m_s/25$ relation. Although, The ratio $m_s/T_{\text{FO}}$ changes with the variation of various portal couplings. 
A large DM mass ($\gtrsim 1$TeV) is required for the phenomena of the dilution of the DM relic density due to an FOPT that occurs around the electroweak scale, since this will ensure that the DM freeze-out temperature is maintained above the temperature at which the FOPT occurs.

In order to obtain the DM abundance and the freeze-out temperature
we first implement the model in $\feynrules$ \cite{Alloul:2013bka} and then
the outputs are fed into $\micromegas$~\cite{Belanger:2006is}.
Note that in the situation where an FOPT occurs around the electroweak scale in the scalar potential and the DM freeze-out temperature is larger than the phase transition temperature, the DM relic density and freeze-out temperature that are estimated using $\micromegas$ no longer remain valid. Such a scenario may appear 
for heavier DM ($\gtrsim 1$~TeV).
However, such a computation is outside the scope of this study, we preserve it for future work.
The present work aims to quantify the dilution factor of the DM relic density caused by the FOEWPT due to the release of entropy, which is discussed in the next subsection. 
\subsection{Estimation of the dilution factor}
\label{dilutionfactor}
In this subsection, we estimate the dilution factor of the DM relic density due to an FOPT along the $h_d$-field direction.
The triplet field values, $h_t$, at the various minima of the effective potential at finite temperature always remain tiny in the region of parameter space that we study for this work.
An FOPT can occur in the early Universe along the singlet-field direction, i.e., along the $h_s$-field direction. 
Such a phase transition around the temperature of the electroweak scale demands a relatively low DM mass~\cite{Ghosh:2022fzp}.
On the other hand, the dilution  effect caused by the FOPT around the electroweak scale on the DM relic density becomes significant only when the DM mass is relatively large $\gtrsim$ 1~TeV. 
Such a phase transition is unlikely to occur for such a heavy DM.
On the contrary, an FOPT around the electroweak scale can happen along the $h_d$-field direction and such a phase transition could dilute the relic density of a heavy DM. 
Therefore, in the context of this work, the relevant FOPT is the transition that happens only along the $h_d$-field direction. 

\begin{figure}[t!]
\begin{center}
\hskip-18pt
\includegraphics[height=5.9cm,width=0.51\linewidth]{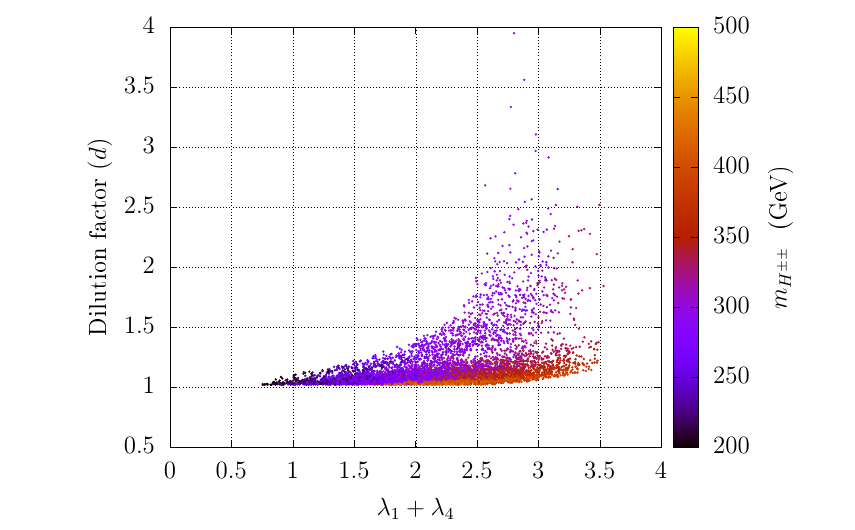}
\hskip 2pt
\includegraphics[height=5.9cm,width=0.51\linewidth]{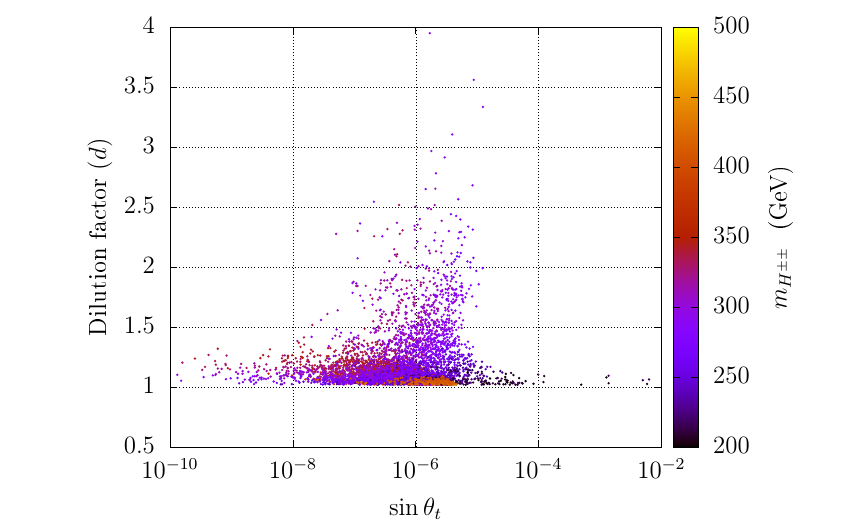}
\hskip 2pt
\includegraphics[height=5.9cm,width=0.51\linewidth]{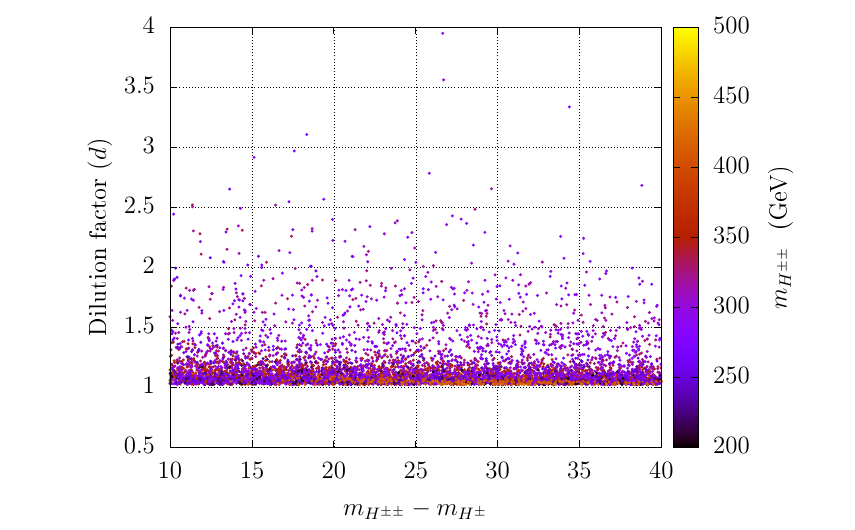}
\caption{Variation of the dilution factor ($d$) with the effective quartic coupling $\lambda_1+\lambda_4$ (top-left), mixing parameter $\sin\theta_t$ (top-right) and the mass split,  $m_{H^{\pm\pm}} - m_{H^{\pm}}$ (bottom). The palette-color indicates $m_{H^{\pm\pm}}$.}
\label{parameterscan}
\end{center}
\end{figure}
The variation of the dilution factor caused by an FOPT, calculated using equation~\ref{dilutioncase2},  with the effective quartic coupling of the potential ($\lambda_1+\lambda_4$) has been shown in the top-left plot of figure~\ref{parameterscan}. The color in the palette indicates the mass of the doubly-charged Higgs. Note that the dilution factor increases with the increase of the effective quartic coupling $\lambda_1+\lambda_4$. The color variation indicates that a decrease in $m_{H^{\pm\pm}}$ enhances the dilution factor for a given value of $\lambda_1+\lambda_4$. 
Thus, the parameter space with
larger effective quartic couplings ($\lambda_1$, $\lambda_4$) and lower $m_{H^{\pm\pm}}$ are preferred for larger dilution factors resulting from an FOPT along the $h_d$-field direction.
Note that a significant amount of parameter space from this region, the decay width of $\hsm$
into a photon pair is more than $2\sigma$
away from the latest LHC limits. However, we still
preserve those points in these plots to understand the behaviour of the dilution factor with
the model parameters.

We present the variation of dilution factor in terms of more physical parameters, i.e.,  the mixing parameter $\sin\theta_t$ (defined in equation~\ref{tanthetat}) and the triplet-like scalar mass in the top-right plot of figure~\ref{parameterscan}.
It can be observed that we focus on a small mixing limit between the $CP$-even SM-like Higgs boson and the triplet-like scalar to evade various experimental constraints that are coming from the study of the discovered SM-like Higgs boson at the LHC. We observe relatively larger dilution factor for $\sin\theta_t$ in between $10^{-7}$to $10^{-5}$.
Another parameter we vary is the mass split among the triplet-like fields, as discussed in Section~\ref{model}. In the bottom plot of Figure~\ref{parameterscan}, we show the variation of the dilution factor with the mass split $\Delta m = m_{H^{\pm\pm}} - m_{H^{\pm}}$~\footnote{As discussed in section~\ref{model}, to avoid constraints coming from EW precision data and various collider searches, we restrict the mass split to the window $10\, \text{GeV} < \Delta m < 40\, \text{GeV}$.}. From these plots, it is evident that the dilution factor does not exhibit any significant dependence on $\Delta m$. Instead, it is primarily influenced by $\lambda_1 + \lambda_4$ and the masses of the triplet-like scalars.

%

%

\begin{figure}[t!]
\begin{center}
\hskip-18pt
\includegraphics[height=5.8cm,width=0.51\linewidth]{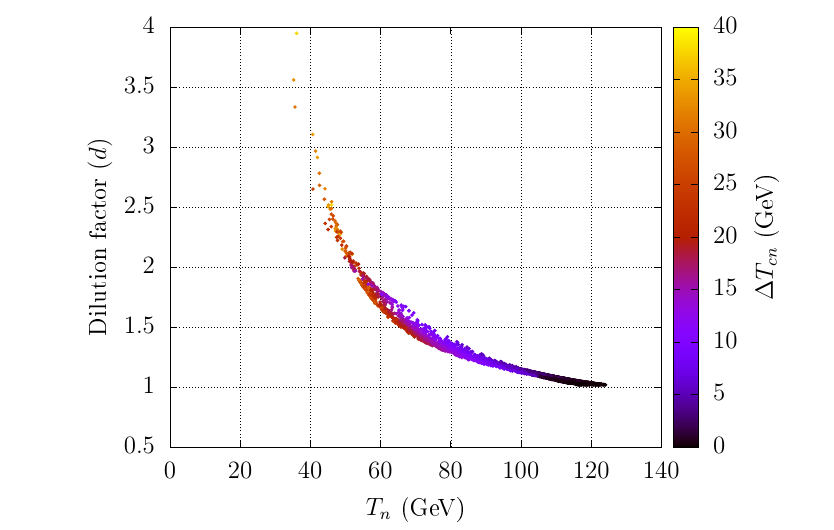}
\hskip 2pt
\includegraphics[height=5.8cm,width=0.51\linewidth]{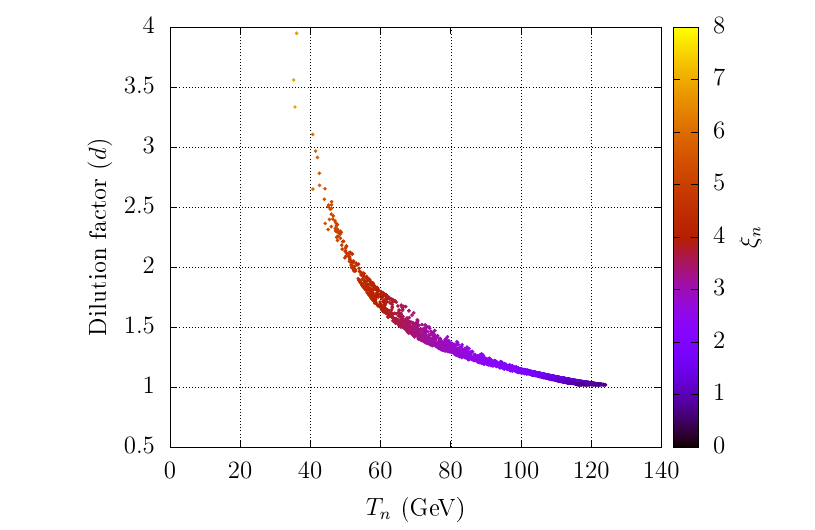}
\caption{Variation of dilution factor ($d$) with the nucleation temperature ($T_n$) of the FOEWPT. The variation of the difference between the critical temperature and the nucleation temperature $\Delta T_{cn}$ and the strength of the phase transition $\xi_n$ are indicated via the palettes in the left and right plots, respectively.}
\label{fig:dilutionplots}
\end{center}
\end{figure}

 One of the most critical parameters of the estimation of dilution is the temperature at which the FOPT occurs. 
 Here, we discuss the correlations between the nucleation temperature ($T_n$) of the phase transition and the dilution factor.
Figure~\ref{fig:dilutionplots} represents the variation of the dilution factor with $T_n$. It indicates that the dilution factor increases with decreasing $T_n$.
The palette color in the left plot indicates the difference between $T_c$ and $T_n$, i.e.,  $\Delta T_{cn}$.
The variation in color indicates that the dilution factor also increases with an increasing $\Delta T_{cn}$. 
In the scenario with a small $\Delta T_{cn}$, the impact of dilution becomes insignificant due to negligible supercooling. As opposed to that, significant supercooling can occur for the large separations between $T_c$ and $T_n$, which enhances the dilution factor. This result is consistent with the findings in references~\cite{Megevand:2007sv, Wainwright:2009mq, Bian:2018mkl, Bian:2018bxr}. The dependence of the supercooling with the dilution factor can be understood from the relation in equation~\ref{dilutioncase2}. In the right plot of figure~\ref{fig:dilutionplots}, the strength of the phase transition ($\xi_n$) is defined as, $\xi_n = \frac{\sqrt{<h_d (T_n)>^2_{\text {BP}} - <h_d (T_n)>^2_{\text {SP}}}}{T_n}$ where `BP' and `SP' represent the broken phase and the symmetric phase, respectively.
$\xi_n$ is indicated by the palette color. The color variation indicates that the dilution factor increases with $\xi_n$. 
It is expected since a stronger FOPT often results in substantial supercooling, which enhances the dilution factor. 

The relic density would not be diluted if the DM freeze-out temperature, $T_{\text{FO}}$, is less than $\sim$ 40~GeV.
We find that the dilution factor can increase up to around $\sim$ 3 at $T_n \sim 40$~GeV.
On the other hand, the effect of dilution is almost negligible for $T_n \gtrsim$ 120~GeV.
Thus, there would always be a considerable dilution of the relic density for any DM candidate that freezes out at temperatures larger than 120 GeV. As a result, the dilution effect would only be significant for significantly heavy DM.

The DMDD and DMID cross-sections are influenced by the relic density of DM in the present Universe. For an underabundant DM scenario, these cross-sections scale down with the factors \(\xi_{\rm DM}\) and \(\xi_{\rm DM}^2\), respectively, as shown in Eqs.~\ref{eq:sigmadd} and \ref{indirectcross}. Consequently, if the dilution factor (\(d\)) is not accounted for in the relic density estimation, the DMDD and DMID cross-sections are reduced by factors of \(d\) and \(d^2\), respectively.
Studying the dilution of relic density is essential, as it can rescue some of the samples which are otherwise excluded by DMDD and DMID experimental constraints. As evident from Eqs.~\ref{eq:sigmadd} and \ref{indirectcross}, \(\sigma_{\rm DD}^{\rm SI} \propto \lambda_{SH}^2 / m_S^4\) and \(\langle \sigma v \rangle_{SS \to X\overline{X}} \propto \lambda_{SH}^2 / m_S^2\), while scaling with \(d\) and \(d^2\), respectively. 
Thus, observing a WIMP-like DM signal in future experiments and estimating the DMDD cross-section would indicate that the \(\lambda_{SH}\) value would be smaller by a factor of \(\sqrt{d}\)
 for a given DM mass under a dilution scenario. Similarly, for DMID, \(\lambda_{SH}\) would be reduced by a factor of \(d\).

In addition to this, such a dilution caused by FOPT can bring down the DM relic density value below the Planck experiment measured value and save the point that would otherwise be excluded due to the overabundance.
The experimentally accessible parameter space for DMDD, in terms of $\lambda_{SH}$ and $m_S$, may fall below the irreducible neutrino background, often referred to as the "neutrino floor/fog"~\cite{Billard:2013qya}. In this case, the dark matter signal would be overwhelmed by the neutrino background, rendering detection extremely challenging for current and near-future DMDD experiments.
 Therefore, such an estimation of the dilution factor due to FOEWPT is crucial for the study of DM phenomenology.
In addition, note that such an FOEWPT can generate a stochastic GW. Thus, in the following subsection, we shall discuss the connection between the dilution factor and the produced GW due to the FOEWPT.
\subsection{Connection between the produced GW and the Dilution factor}
\label{GWdilutioncorrelation}
\begin{figure}[t!]
\begin{center}
\hskip-18pt
\includegraphics[height=6.3cm,width=0.58\linewidth]{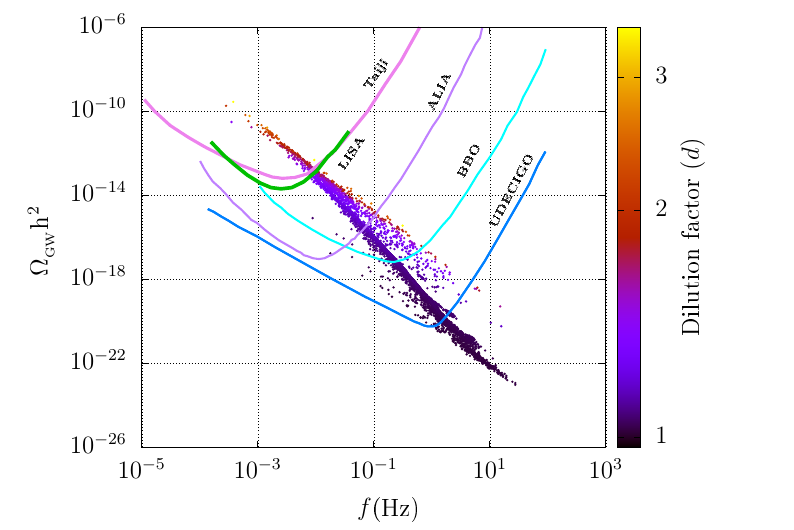}
\caption{Variation of the GW peak amplitude  ($\Omega_{\text{GW}} h^2)_{\text{peak}}$ with the peak frequency ($f_{\text{SW}}$) estimated from the sound wave contribution in the GW power spectrum ($\Omega_{\text{GW}}h^2$) - frequency ($f$) plane. The
palette-color shows the variation of the dilution factor ($d$). The color palette indicates the variation of the dilution factor ($d$). The different colored lines represent the experimental sensitivity curves of future proposed gravitational wave detectors, including LISA, Taiji, aLIGO+, BBO, ALIA, and U-DECIGO.}
\label{fig:gwscanplots}
\end{center}
\end{figure}
Future proposed GW experiments might detect stochastic GW generated during an FOEWPT which also dilutes the relic density of DM.
The dominant contribution of the GW peak intensity among the three mechanisms of the production of GW, as described in section~\ref{dilution}, comes from the sound waves. The details of the estimation of GW intensity, $\Omega_{\text{GW}}h^2$, with frequency ($f$) from the sound waves are presented in Appendix~\ref{GWexpression}~\cite{Ghosh:2022fzp, Chatterjee:2022pxf}.
The GW peak amplitude (($\Omega_{\text{GW}} h^2)_{\text{peak}}$) can be found by putting $f = f_{\text{SW}}$ in equation~\ref{GWsound}. The peak frequency ($f_{\text{SW}}$) is given in equation~\ref{GWsoundfreq}.

In figure~\ref{fig:gwscanplots}, we present the variation of the peak amplitudes  with the peak frequencies of the produced GWs as a result of an FOEWPT in the $\Omega_{\text{GW}}h^2-f$ plane.
The dilution factor of the DM relic density caused by an FOEWPT is presented by the palette-color.
Note that various future proposed space/ground-based GW experiments, such as LISA~\cite{LISA:2017pwj}, ALIA~\cite{Gong:2014mca}, Taiji~\cite{Hu:2017mde}, TianQin~\cite{TianQin:2015yph}, aLigo+~\cite{Harry:2010zz}, Big Bang Observer (BBO)~\cite{Corbin:2005ny} and  Ultimate(U)-DECIGO~\cite{Kudoh:2005as}, etc., have different sensitivity regions across the peak intensity and peak frequency ranges, as illustrated in the figure. Therefore, the study of the variation of the peak frequency and the peak amplitude of the produced GW is crucial for detecting the spectrum in future proposed GW experiments.
The color variation in figure~\ref{fig:gwscanplots} indicates that the dilution factor increases as both GW peak amplitude and peak frequency increase.
Such a connection between GW intensity and dilution factor can be realized on their dependence on the strength of the phase transition, $\xi_n$.
Thus, a stronger FOEWPT forecasts a bigger GW peak amplitude as well as a larger dilution factor. Therefore, observing the GW peak amplitude and the peak frequency in some of the future proposed GW experiments can reveal the range of the DM dilution factor of the DM relic density due to a release of entropy during the FOEWPT.
\subsection{Probing the parameter space}
In section~\ref{model}, we describe the chosen parameter space for the model, with a particular focus on the triplet sector and a moderate mass splitting of $10, \text{GeV} < \Delta m < 40, \text{GeV}$ to satisfy constraints from LHC searches~\cite{Ashanujjaman:2021txz}. In this range, charged Higgses primarily decay into neutral ones and off-shell $W$-bosons, producing soft leptons or jets that are challenging to reconstruct at the LHC. Similarly, neutral triplet-like Higgses decay into neutrinos or such as $b\bar{b}$, $t\bar{t}$, $ZZ$, $Zh$, and $hh$, resulting in final states dominated by SM backgrounds. Consequently, the production of $H^{\pm\pm} H^{\mp\mp}$ and $H^{++} H^{--}$ pairs remains elusive at the LHC, leaving this parameter space largely unconstrained. However, future lepton colliders, with their cleaner experimental environments, present promising opportunities to probe this parameter space~\cite{Ashanujjaman:2022tdn}. A recent work indicates that $e^{+}e^{-}$ colliders with center-of-mass energies of 500~GeV and 1~TeV could achieve 5$\sigma$ sensitivity over a substantial portion of this parameter space~\cite{Ashanujjaman:2022tdn}. Further dedicated analyses, including those involving other proposed lepton colliders such as the muon collider~\cite{Accettura:2023ked}, could refine these prospects by directly targeting the parameter space explored in this work. We plan to study this in our future work. In addition, a substantial portion of the parameter space can be explored by investigating the di-photon decays of $\hsm$ at the LHC.

Future proposed GWs experiments are expected to probe specific regions of the parameter space, as discussed in the previous section. A comprehensive analysis in this direction can be found in Ref.~\cite{Ghosh:2022fzp}. It is worth noting that if a GWs signal corresponding to an FOPT near the electroweak scale is detected in the near future, and concurrent evidence for TeV-scale DM emerges from the DM direct and/or indirect detection experiments, the significance of the present work would be greatly enhanced.
However, to refine our understanding of the dilution factor and its correlation with GW signals, several theoretical improvements are necessary. Specifically, the modeling of GW production from FOPTs requires enhanced precision, including a more accurate description of bubble dynamics, the contributions of sound waves, and magnetohydrodynamic turbulence in the plasma. Moreover, a robust determination of the effective potential at finite temperatures is essential to better characterize the phase transition dynamics and its associated phenomenology. These developments, coupled with data from next-generation GWs detectors like LISA, Taiji, BBO, ALIA and U-DECIGO, could provide deeper insights into the interplay between FOPTs and DM, providing a robust framework to test BSM scenarios like the one considered in this work.

\section{Conclusions}
\label{conclusions}
Despite many successes of the SM, it cannot explain 
the origin of the observed baryon asymmetry of the Universe, the non-vanishing masses of neutrinos and accommodate a DM candidate. 
In this article, we consider the $Z_3$-symmetric complex singlet scalar extended type-II seesaw model that can simultaneously provide solutions to these issues within a single framework.
An SFOEWPT is one of the essential requirements for triggering EWBG that can generate the observed BAU at the electroweak scale.
Such an FOEWPT could dilute the relic density of DM due to an injection of entropy and a release of latent heat. 

In the current model, the triplet sector can modify the SM Higgs potential in favor of an FOEWPT along the $h_d$-field direction even when the singlet scalar fields are heavy.
 As a result, the singlet scalar DM can get heavier while still allowing for an FOEWPT in the scalar potential. In such a scenario, a significant dilution in the DM relic abundance is possible for a heavier DM that freezes out much before the FOEWPT takes place.

We find the regions of parameter space of the triplet sector of the scenario that corresponds to an FOEWPT and can generate a sizable amount of dilution. 
Such a region prefers  relatively light  $m_{H^{\pm\pm}}$ and larger effective quartic couplings $\lambda_1+\lambda_4$.
In the case of supercooling, the released entropy caused by the FOEWPT could dilute the DM relic density up to a factor of around 3 in our scenario.
Furthermore, we show the variation of the dilution factor with the temperature at which the phase transition occurs. We note that the dilution factor increases with decreasing $T_n$.
The difference between $T_c$ and $T_n$ is also an important factor in this estimation.
A larger gap between $T_c$ and $T_n$ leads to a significant amount of supercooling during the phase transition which, in turn, enhances the dilution factor.
DM relic density dilution starts only for $T_n \lesssim 120$~ GeV. Thus, the DM with a freeze-out temperature much larger than this, which is possible in this model, could experience significant dilution in its relic abundance.
Such a dilution might retrieve some of the regions of parameter space that
were previously ruled out by the measured value of the DM relic density and/or the latest
constraints from the DMDD experiments.
On the contrary, there will be no dilution in the abundance of DM whose freeze-out temperature is $\lesssim$ 40 GeV and such a scenario corresponds to a relatively light DM ($\lesssim$ 1 TeV).

Dilution of the DM relic density is only possible in this model when the DM is heavy
and an FOEWPT occurs in the early Universe. The type-II seesaw sector of this current
model that can alter the SM-like Higgs potential in favor of an FOEWPT would get tested
by studying the di-photon decays of the SM-like Higgs boson at the HL-LHC. A null-finding
of new physics at the HL-LHC would rule out the possibility of an FOEWPT triggered by
the triplet sector and hence the possibility of dilution of relic density of a TeV-scale DM in
such a scenario.

The dilution factor increases with the strength of the FOEWPT. In addition, stochastic GW can originate from such an FOEWPT in the early Universe.
Thus, a connection between the dilution factor and the generation of the stochastic GW, as a result of an FOEWPT, is discussed in this work.
The dilution factor increases with GW peak amplitude and peak frequency.
Therefore, observing peak amplitude and peak frequency of GW produced from an FOEWPT in some of the future proposed space/ground-based GW experiments
can reveal the range of the DM dilution factor of the DM relic density due to the
release of entropy during the same phase transition.

\section*{Acknowledgments}
This work was initially carried out at the Harish-Chandra Research Institute (HRI), where SR was supported by the Department of Atomic Energy (DAE), Government of India, through the Regional Centre for Accelerator-based Particle Physics (RECAPP) at HRI. SR was also supported by the Excellence in Research Award from Infosys. The revised version of this work was completed at Argonne National Laboratory, where SR is supported by the U.S.~Department of Energy under contract No.~DEAC02-06CH11357. SR would like to thank the University of Chicago and Fermilab, where significant portions of this work were undertaken. SR also expresses gratitude to Asesh Krishna Datta and Yang Zhang for insightful discussions. Additionally, SR acknowledges the use of high-performance scientific computing facilities at HRI and RECAPP.
%
\appendix
\label{appendix}
\section*{Appendix}
\section{Contribution of sound waves to gravitational waves power spectrum}
\label{GWexpression}
The dominant contribution to the total GW power spectrum coming from the sound waves, $\Omega_{\text{sw}} {\rm h^2}$, is given by~\cite{Hindmarsh:2013xza, Giblin:2013kea, Giblin:2014qia, Hindmarsh:2015qta}
\begin{equation}\label{GWsound}
\Omega_{\text{SW}}{\rm h}^2=2.65\times 10^{-6} \Upsilon(\tau_{SW}) \left(\dfrac{\beta}{H_{\star}} \right) ^{-1} v_{w} \left(\dfrac{\kappa_{v} \alpha}{1+\alpha}\right)^2 \left(\dfrac{g_*}{100}\right)^{-\frac{1}{3}}\left(\dfrac{f}{f_{\text{SW}}}\right)^{3} \left[\dfrac{7}{4+3 \left(\dfrac{f}{f_{\text{SW}}}\right)^{2}}\right]^{\frac{7}{2}}\,\,,
\end{equation}
where $f_{\text{SW}}$ is the present day peak frequency and is given by
\begin{equation}\label{GWsoundfreq}
f_{\text{SW}}=1.9\times10^{-5}\hspace{1mm} \text{Hz} \left( \dfrac{1}{v_{w}}\right)\left(\dfrac{\beta}{H_{\star}} \right) \left(\dfrac{T_n}{100 \hspace{1mm} \text{GeV}} \right) \left(\dfrac{g_*}{100}\right)^{\frac{1}{6}}\,\,.
\end{equation}
In the above relations, $H_{\star}$ is the Hubble rate just after the end of GW production and $T_{\star}$ is the temperature at that time.
For this work, we consider $T_{\star} \approx T_n$. The dimensionless quantity that is related to the energy budget of the phase transition, $\alpha$, is defined as~\cite{Espinosa:2010hh}
\beq
\alpha = \frac{\rho_{\text{vac}}}{\rho^*_{\text{rad}}} = \frac{1}{\rho^*_{\text{rad}}}\left[T\frac{\d \Delta V(T)}{\d T} - \Delta V(T)\right]\Bigg|_{T_*}.
\eeq
The parameter $\beta$ that estimates the inverse time duration of the phase transition is given by
\beq
\beta = -\frac{d S_3}{dt}\Bigr|_{t_*} \simeq \dfrac{\dot{\Gamma}}{\Gamma} = H_*T_* \frac{d(S_3/T)}{dT}\Bigr|_{T_*}.
\eeq
The parameter $\kappa_v$ estimates the fraction of latent heat energy transferred into the bulk motion of the fluid and is given by~\cite{Espinosa:2010hh}
\beq
\label{eq:76}
\kappa_v \simeq \left[ \dfrac{\alpha}{0.73+0.083\sqrt{\alpha}+\alpha}\right]\,\,.
\eeq
The suppression factor 
$\Upsilon(\tau_{SW})$ appears due to the finite lifetime of sound waves and is given by~\cite{Guo:2020grp, Hindmarsh:2020hop}
\beq
\Upsilon(\tau_{SW}) = 1 - \frac{1}{\sqrt{1 + 2 \tau_{\text{sw}} H_{\ast}}} \, \, ,
\label{eq:upsilon}
\eeq
Where $\tau_{\text{sw}}$ is the lifetime of the sound waves. It can be estimated from the timescale in which the turbulence develops. It is approximately given by~\cite{Pen:2015qta,Hindmarsh:2017gnf}:
\begin{eqnarray}
\tau_{\text{sw}} \sim \frac{R_{\ast}}{\bar{U}_f} \, \, ,
\end{eqnarray}
where $R_{\ast}$ denotes the mean bubble separation related to the phase transition duration. It is given by $R_{\ast} = (8\pi)^{1/3} v_w /\beta$~\cite{Hindmarsh:2019phv, Guo:2020grp}. The quantity $\bar{U}_f$ represents the root-mean-squared (RMS) velocity and it is given by $\bar{U}_f = \sqrt{3 \kappa_v \alpha/4}$~\cite{Hindmarsh:2019phv,Weir:2017wfa}. More details related to the computation of the GW power spectrum from the FOPT in the present model have been discussed in reference~\cite{Ghosh:2022fzp}.
%

\end{document}